\documentclass[conference,10.5pt]{IEEEtran}
\IEEEoverridecommandlockouts
\usepackage{cite}
\usepackage{amsmath,amssymb,amsfonts}
\usepackage{algorithmic}
\usepackage{graphicx}
\usepackage{color}
\usepackage{subfig}
\usepackage{stackengine}
\usepackage{comment}
\usepackage{ulem}
\usepackage{tabularx}
\usepackage{bm}
\usepackage{bbold}
\usepackage[utf8]{inputenc}

\usepackage{mathtools}
\usepackage[singlespacing]{setspace} 

\usepackage{lettrine}
\usepackage[font=scriptsize]{caption}

\usepackage[colorlinks, citecolor=blue]{hyperref} 

\usepackage[flushleft]{threeparttable} 
\usepackage{booktabs}

\usepackage{textcomp}
\usepackage{xcolor}

\usepackage[linesnumbered, ruled]{algorithm2e}
\SetKwRepeat{Do}{do}{while}%

\SetCommentSty{mycommfont}

\def\BibTeX{{\rm B\kern-.05em{\sc i\kern-.025em b}\kern-.08em
    T\kern-.1667em\lower.7ex\hbox{E}\kern-.125emX}}
    
\usepackage{xspace}
\usepackage{ifthen}
\usepackage[inline]{enumitem}

\newboolean{showcomments}
\setboolean{showcomments}{true}
\ifthenelse{\boolean{showcomments}}
{
}

\usepackage{subfig}
\usepackage{fancyhdr}

\newcommand*{\affmark}[1][*]{\textsuperscript{#1}}

\usepackage{fancyhdr}
\begin{document}

\title{\huge RL meets Multi-Link Operation in IEEE 802.11be: Multi-Headed Recurrent Soft-Actor Critic-based Traffic Allocation}

\author{\IEEEauthorblockN{ Pedro Enrique Iturria-Rivera\affmark[1], \IEEEmembership{Graduate Student Member,~IEEE}, Marcel~Chenier\affmark[2], Bernard~Herscovici\affmark[2],\\ Burak~Kantarci\affmark[1], \IEEEmembership{Senior Member,~IEEE} and Melike Erol-Kantarci\affmark[1], \IEEEmembership{Senior Member,~IEEE}}
\IEEEauthorblockA{\affmark[1]\textit{School of Electrical Engineering and Computer Science, University of Ottawa, Ottawa, Canada}}  \affmark[2]\textit{NetExperience., Ottawa, Canada}\\
Emails:\{pitur008, burak.kantarci, melike.erolkantarci\}@uottawa.ca,  \{marcel, bernard\}@netexperience.com
}



\maketitle
\begin{abstract}
\textbf{IEEE 802.11be ---Extremely High Throughput---, commercially known as Wireless-Fidelity (Wi-Fi) 7 is the newest IEEE 802.11 amendment that comes to address the increasingly throughput hungry services such as Ultra High Definition (4K/8K) Video and Virtual/Augmented Reality (VR/AR). To do so, IEEE 802.11be presents a set of novel features that will boost the Wi-Fi technology to its edge. Among them, Multi-Link Operation (MLO) devices are anticipated to become a reality, leaving Single-Link Operation (SLO) Wi-Fi in the past. To achieve superior throughput and very low latency, a careful design approach must be taken, on how the incoming traffic is distributed in MLO capable devices. In this paper, we present a Reinforcement Learning (RL) algorithm named Multi-Headed Recurrent Soft-Actor Critic (MH-RSAC) to distribute incoming traffic in 802.11be MLO capable networks. Moreover, we compare our results with two non-RL baselines previously proposed in the literature named: Single Link Less Congested Interface (SLCI) and Multi-Link Congestion-aware Load balancing at flow arrivals (MCAA). Simulation results reveal that the MH-RSAC algorithm is able to obtain gains in terms of Throughput Drop Ratio (TDR) up to $35.2\%$ and $6\%$ when compared with the SLCI and MCAA algorithms, respectively. Finally, we observed that our scheme is able to respond more efficiently to high throughput and dynamic traffic such as VR and Web Browsing (WB) when compared with the baselines. Results showed an improvement of the MH-RSAC scheme in terms of Flow Satisfaction (FS) of up to $25.6\%$ and $6\%$ over the the SCLI and MCAA algorithms.}
\end{abstract}


\section{Introduction}

\lettrine[findent=1pt]{\textbf{W}}{ }ireless Fidelity (Wi-Fi) technology has experienced a rapid evolution in the last four years with the introduction of Wi-Fi $6$ and Wi-Fi $6$E in $2019$ and $2021$, respectively. This technological development momentum seems to keep going and IEEE 802.11be ---Extremely High Throughput (EHT)--- known commercially as Wi-Fi $7$ is expected to become reality by $2024$. Even more, IEEE 802.11 has started in $2022$ the first conversations about devising the next generation of Wi-Fi (named Wi-Fi 8) believed to be released by $2028$ \cite{Choi}.
The still under development IEEE 802.11be amendment is considered with the goal of dealing with highly congested network environments with stringent requirements in terms of high throughput and low latency. To do so, the amendment proposes to increase the channel bandwidth up to $320$ MHz, a higher modulation rate up to 4096 QAM and different than all other Wi-Fi generations, Wi-Fi 7 introduces Multi-Link Operation (MLO) and Multiple Resource Units (MRU) capabilities.

\begin{figure*}[h]
\center
  \includegraphics[scale=0.61]{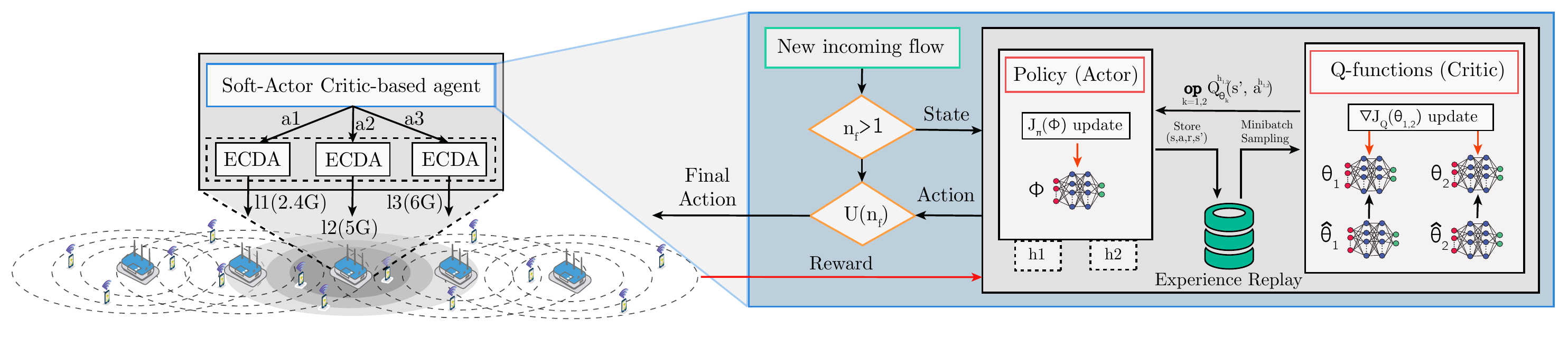}
  \setlength{\belowcaptionskip}{-5pt}
  \caption{Overview of the MH-RSAC based traffic allocation policy in IEEE 802.11be MLO: Upon an incoming flow the agent will decide the percentage (a1, a2, a3) of traffic flow allocated to each available interface (l1, l2, l3) based on the observed state.} 
  \label{system_overview}
\end{figure*}

In this work, we focus our interest towards MLO in 802.11be. Specifically, MLO allows Multi-Link Devices (MLD)s to concurrently use their available interfaces for multi-link communications. In this context, we intend to optimize the traffic allocation policy over the available interfaces with the aid of Reinforcement Learning (RL). RL has proven its effectiveness to deal with challenging problems in wireless networks\cite{9700667}. Thus, in this paper we propose to utilize a Soft-Actor Critic (SAC) algorithm named Multi-Headed Recurrent SAC (MH-RSAC). We choose the SAC algorithm over other Actor-Critic methods such as \textit{Advantage Actor Critic} (A2C) or \textit{Proximal Policy Optimization} (PPO) due its consistent performance in many RL challenging tasks \cite{Haarnoja2018}. Differently from others, the SAC algorithm maximizes its rewards altogether with the entropy which benefits the agent's exploration. We propose for the first time, to the best of our knowledge, a novel SAC-based traffic-to-link allocation policy in 802.11be MLO capable networks. In Fig. \ref{system_overview}, we show the scenario and system overview of our proposed scheme. Here, each agent residing in the Access Point (AP) equipment, decides the incoming traffic distribution percentage (a1, a2, a3) according to the MLO capabilities. For instance, if the number of interfaces of the end station corresponds to $n_f>1$, the agent will propose a set of actions for the case of two and three interfaces, respectively. Finally, the function $U(n_f)$ selects the final action based on the actual interface number. Note that, when $n_f=1$ the agent is not involved since all the traffic is passed to the only available interface. This design allows to have one agent per AP instead of two. In addition, we consider the non-Markovian behavior of the scenario and include a recurrent neural network in the model design. Moreover, we improve even further the performance of our agent by utilizing a modified reward function named \textit{rewards with hindsight} that consider the goals of the baselines. Finally, we utilize in our simulations, three traffic flow types: Web Browsing (WB), High Definition (HD)/ Ultra High Definition (4K) and Virtual Reality (VR). For the latter, we derive the VR CDFs based on \cite{9685808} and present the given equations to be utilized in any wireless simulator.

We compare our results with two non-machine learning type baselines presented in \cite{Lopez-Raventos2021} named \textit{Single Link Less Congested Interface} (SLCI), \textit{Multi Link Congestion-aware Load balancing at flow arrivals} (MCAA), respectively. Our results show an improvement in terms of Throughput Drop Ratio (TDR) with a gain of  up to $35.2\%$ and $6\%$ when compared with SLCI, MCAA non-RL baselines, respectively. TDR is defined as the percentage of traffic in terms of throughput dropped after a traffic allocation decision.  Finally, we observed that our scheme is able to respond more efficiently to high throughput and dynamic traffic such as VR and Web Browsing (WB) with  an improvement in terms of Flow Satisfaction (FS) up to $25.6\%$ and $6\%$ for the SCLI and MCAA, respectively. 

The rest of this paper is organized as follows. Section \ref{Section2} presents the existing works related to MLO in 802.11be. System model is demonstrated in Section \ref{section3}.
 Section \ref{section4} provides a description of the MH-RSAC scheme and the considerations taken on its design, including the Markov Decision Process (MDP) and the description of the baselines. Section \ref{section5} depicts the traffic considerations in this work. Section \ref{Section6} presents our proposed scheme's performance evaluation and comparison with the baselines. Finally, section \ref{Section7} concludes the paper.

\section{Related work}\label{Section2}
To the best of our knowledge, this is the first work that propose a Reinforcement Learning-based traffic allocation policy in 802.11be Multi-Link Operation. However, 802.11be MLO has been recently of great interest among researchers. Some of these works are referred bellow.

In \cite{Lacalle2021}, the authors present a study of MLO in Wi-Fi 802.11be in the context of a sub-standard of 802.1, Time Sensitive Networking (TSN). Results show that MLO is able to improve latency over a single link, however the usage of uncorrelated channels are recommended to increase the average SNR perceived by the stations. In addition, the authors in \cite{Carrascosa3} perform an experimental study using a real dataset measurement; a dataset containing 5 GHz spectrum occupancy measurements. In the before mentioned study, it is proved that when MLO is available in symmetrical occupied channels, latency can be reduced by one order of magnitude over single link operation. Interestingly, it is observed that when the channels do not fulfill the symmetry condition, the latency performance worsens. The authors solve the previous issue by adding a backoff counter in each interface to avoid packet collision. Related to our current work, the authors in \cite{Lopez-Raventos2021} propose three non-dynamic flow traffic allocation policies in Wi-Fi 802.11be: SLCI, MCAA and \textit{Multi Link Same Load to All interfaces} (MLSA), respectively. They prove that MLO performance is tightly related to the efficiency of the flow distribution over the available interfaces in a Multi-Link Device. The authors present in \cite{lopez2}, a dynamic flow traffic allocation policy that is able to adapt its traffic policy not only upon the arrival of the traffic flow but every $1$ second. Finally, surveys can be found throughout the literature that delve into the challenges and perspective of 802.11be and MLO such as in \cite{Yang2020,Carrascosa1}.


 


\section {System Model} \label{section3}

In this work, we utilize an IEEE 802.11be network with a predefined set of $\mathcal{M}$ APs and $\mathcal{N}$ stations attached per AP. In addition, we consider all APs to be MLO-capable with a number of available interfaces $n_f=3$ whereas in the case of the stations the MLO capability can vary. This design decision is taken based on the fact that single device and multi-device terminals will coexist in real scenarios due to terminal manufacturer diversity. Furthermore, stations are positioned in two ways with respect their attached AP: $80\%$ of the users are positioned randomly in a radius $r\sim[1-8] $ m and the rest with a radius $r\sim[1-3]$ m. All APs and stations support up to a maximum of 16-SU multiple-input multiple-output (MIMO) spatial streams. The path loss model corresponds to the enterprise model described in \cite{IEEEP802.112015}. In addition, we consider an adaptive control rate based on Signal to Noise Ratio (SNR) with a maximum 4096 QAM modulation rate.

\section{Multi-Headed Recurrent Soft-Actor Critic} \label{section4}
In the current section we discuss the details and considerations taken in the design of the Multi-Headed Recurrent Soft-Actor Critic (MH-RSAC) agent. 

\subsection{Soft-Actor Critic}\label{soft-actor}
The SAC agent introduced initially in \cite{Haarnoja2018} is a maximum entropy, model-free and off-policy actor-critic method that outperformed most of the \textit{state-of-the-art} RL algorithms such as \textit{Twin Delayed Deep Deterministic Policy Gradient} (TD3) and PPO. The success of SAC relays on the inclusion of a policy entropy term into the reward function to encourage exploration. In addition, it reutilizes the successful experience of the minimum operator in the selection of the double Q-Functions that comprise the critic in the \textit{Deep Deterministic Policy Gradient} (DDPG) algorithm \cite{Fujimoto2018}. 

In this work, we utilize a discrete SAC agent presented in \cite{Christodoulou} that provides an adaptation of the original SAC agent to the discrete action space. Moreover, we take into consideration the results obtained recently in \cite{Zhou22}. In such work, the authors study the discrete SAC and find that such algorithm is affected by Q-Value underestimation due to the usage of the minimum operator. In consequence, they propose to substitute the minimum operator by an average operator that allowed to reduce the bias of the lower bound of the double critics. In addition to the previous considerations we modify the structure of the critic and the actor of the discrete SAC to allow multi-output. To do so, we diverge the model in two heads as indicated by Fig. \ref{net}: one head (Head 1) producing the actions when two interfaces are available at the station end and another (Head 2) when three are available. The reason behind this design decision corresponds to the practical assumption, briefly mentioned in Section \ref{section3}, that terminals with diverse MLO capabilities will coexist. Thus, instead of utilizing two agents handling each individual case, we propose to utilize one agent capable of deciding in both circumstances. In such way, we can reduce the computational load required to train and run two agents simultaneously.

\normalem 
\begin{algorithm}

\algsetup{linenosize=\tiny}
 \scriptsize

Initialize actor's policy $\phi$, critic's Q-function parameters $\theta_1$, $\theta_2$ and experience replay buffer $\mathcal{D}$. Set critic's Q-target function equal to main parameters $\hat{\theta}_1 \leftarrow \theta_1$ and $\hat{\theta}_2 \leftarrow \theta_2$. $\bm{op} \in  \{\texttt{min, avg}\}$ corresponds to the operator, $L_{per}$ the update periodicity, $n_{up}$ number of updates per learning step and $\rho \in [0,1]$ is the polyak factor. \\
\For{environment step $t\gets1$ \textbf{to} $T$}{ 

    Observe state $s = [C_{1}^o, C_{2}^o, C_{3}^o, O_{f}, T_{id}]$ and select action for each head $a^{h_{1,2}} \sim \pi_{\phi}^{h_{1,2}}(\cdot|s)$\\
    Execute $a$ in the environment\\
    Observe next state $s'$ and reward $r$\\
    Store $(s, a, r , s')$ in experience replay buffer $\mathcal{D}$\\
    \If{$t$ \textbf{mod} $L_{per} = 0$}{
       \For{$n \gets 1$ to $n_{up}$ }{ 
            Randomly sample a batch of transitions with size $B$ from $\mathcal{D}$\\
            Compute targets for the Q-functions (Critic) for each head $h_{1,2}$: \\
            where $\Tilde{a}'^{h_{1,2}} \sim \pi_{\phi}^{h_{1,2}}(\cdot|s')$\\
            
            Update Q-functions (Critic) considering each head  $h_{1,2}$:\\
            $L_1^{h_{1,2}}(\theta_1,\mathcal{D}) =  E_{(s,a,r,s') \in B}[(Q^{h_{1,2}}_{\theta_1}(s,a^{h_{1,2}}) - y_1(r, s'))^2]$ \\
            
            $L_2^{h_{1,2}}(\theta_2,\mathcal{D}) = E_{(s,a,r,s') \in B}[(Q^{h_{1,2}}_{\theta_2}(s,a^{h_{1,2}}) - y_2(r, s'))^2]$ \\
            
            $L(\theta^{h_{1,2}}, \mathcal{D}) = L_1^{h_{1,2}}(\theta_1,\mathcal{D}) + L^{h_{1,2}}_2(\theta_2,\mathcal{D})$\\
            Update policy (Actor) considering each head $h_{1,2}$:\\
            
            $L^{h_{1,2}}_1(\phi^{h_1},\mathcal{D}) =  E_{(s) \in B}[\bm{op}_{k=1,2}Q_{\hat{\theta}_1,k}^{h_{1,2}}(s,\Tilde{a}'^{h_{1,2}}) - \alpha\log\pi_{\phi}^{h_{1,2}}(\Tilde{a}'^{h_{1,2}}|s')]$ \\
            
            $L^{h_{1,2}}_2(\phi^{h_2},\mathcal{D}) = E_{(s) \in B}[\bm{op}_{k=1,2}Q_{\hat{\theta}_1,k}^{h_{1,2}}(s,\Tilde{a}'^{h_{1,2}}) - \alpha\log\pi_{\phi}^{h_{1,2}}(\Tilde{a}'^{h_{1,2}}|s')]$ \\
            
            $L(\phi^{h_{1,2}}, \mathcal{D}) = L^{h_{1,2}}_1(\phi^{h_1},\mathcal{D}) + L^{h_{1,2}}_2(\phi^{h_2},\mathcal{D})$\\
            
            Finally, update Q-functions target networks with Polyak averaging:\\
            $\hat{\theta}_i \leftarrow \rho\hat{\theta}_i + (1 - \rho)\theta_i$ for $i = 1,2$       
       }
    }
} 
\caption{Multi-Headed Recurrent Soft-Actor Critic}
 \medskip
 \label{soft_algo}
\end{algorithm}
\subsection{Dealing with Non-Markovian environments in RL}
Reinforcement Learning faces some challenges when dealing with non-Markovian environments. The reason is that RL main goal is to maximize the reward $R_{t+1}$ given a certain state $S_t$  and action $A_t$ as: $R= E[R_{t+1}| S_t, A_t]$. When the previous relationship is not fulfilled due partial observability of the Markovian state, the observation state should include more that one input and utilize a portion of the interaction history\cite{Hausknecht2015,Li15}. 
\begin{figure}[h]
\center
  \includegraphics[scale=0.65]{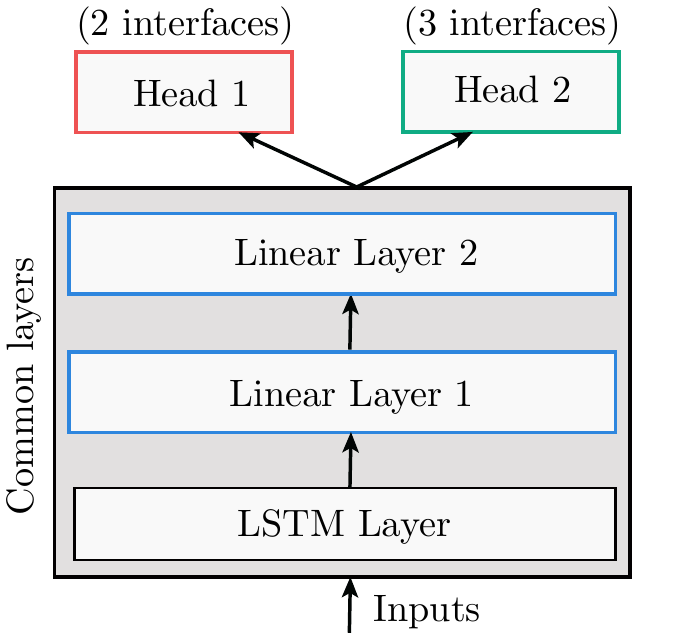}
  \setlength{\belowcaptionskip}{-5pt}
  \caption{Network structure used in the Multi-Headed Recurrent Soft-Actor Critic agent.} 
  \label{net}
\end{figure}

\begin{figure}[h]
\center
  \includegraphics[scale=0.75]{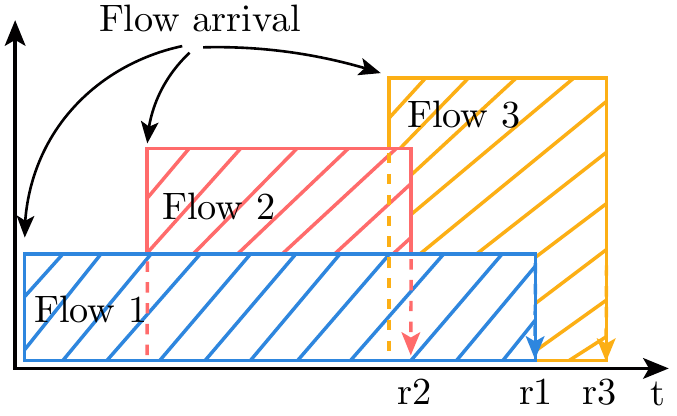}
  \setlength{\belowcaptionskip}{-1pt}
  \caption{Non-Markovian nature of the scenario: Rewards are not only dependant on the current state but affected by other arrival flows.} 
  \label{non-markov}
\end{figure}

In Fig. \ref{non-markov}, we show an example of the non-Markovian behavior of our environment. For instance, the reward obtained upon the arrival of Flow $1$, $r1$ is affected by the arrival of Flow $2$. That means that the reward obtained given the action taken upon Flow $1$ arrival is affected as well by the action taken upon Flow $2$ arrival as well. Such behavior violates the assumption of being on the presence of an MDP. To solve the beformentioned issue, we include as the first layer of the model's structure of the MH-RSAC a \textit{Long Short-Term Memory} (LSTM) neural network as shown in Fig. \ref{net}. In addition, we add two more linear layers after the LSTM layer.

\subsection{State space selection}
As discussed in previous subsections, the environment is modeled as a Partially Observable Markov Decision Process (POMDP), thus a sequence of the observation space is considered instead one instance. The action space is defined as:

\begin{equation}
\begin{split}
    \bm{s}_t = [\{C_{1,t-W}^o, C_{2,t-W}^o, C_{3,t-W}^o, O_{f,t-W}, T_{id,t-W}\},\\\{C_{1,t-W-1}^o, C_{2,t-W-1}^o, C_{3,t-W-1}^o, O_{f,t-W-1},\\ T_{id,t-W-1}\},...,\{C_{1,t}^o, C_{2,t}^o, C_{3,t}^o, O_{f,t}, T_{id,t}\}],
    \end{split}
\end{equation}

where $W$ corresponds to the size of the interaction window, $C_{1}^o, C_{2}^o, C_{3}$ the occupancy of $2.4$ GHz, $5$ GHz and $6$ GHz interface, respectively,  $O_{f}$ the ratio of active flows and $T_{id}$ corresponds to the type of the upcoming flow. Furthermore, we define $O_{f}$ and $T_{id}$, respectively as:
\begin{equation}
    O_{f} = A_f/N_A, 
\end{equation}
where $A_f$ corresponds to the number stations receiving traffic flows and $N_A$ the number of stations attached to the corresponding AP. 
\begin{equation}
     T_{id} =
    \begin{cases}
     0.33 & \text{if } \textit{UHD/4K traffic},\\
     0.66 & \text{if } \textit{VR traffic},\\
     1 & \text{else } \textit{WB traffic}
   \end{cases}
    \label{type_traffic}
\end{equation}

\subsection{Action space selection}
In this subsection, we proceed to describe the action space of the MH-RSAC scheme. As discussed in subsection \ref{soft-actor}, the MH-RSAC structure is comprised by two heads: one providing the action output when two interfaces are available and another for the case when three are. Thus, the action space can be defined as:

\begin{align}
\bm{a}_t =
 \begin{cases}
         \bm{a}_{h1,t} = [\{a_{1,t-W}, a_{2,t-W}\},\\\{a_{1,t-W-1}, a_{2,t-W-1}\}\\,..., \{a_{1,t}, a_{2,t}\}], & \text{if } n_f = 2,\\
         \bm{a}_{h2,t} = [\{a_{1,t-W}, a_{2,t-W}, a_{3,t-W}\},\\ \{a_{1,t-W-1}, a_{2,t-W-1}, a_{3,t-W-1}\}\\,..., \{a_{1,t}, a_{2,t}, a_{3,t}\}], & \text{if } n_f = 3
\end{cases}
\end{align}

where $\bm{a}_{k \in \{1,2,3\}}$ refers to the fraction of the total flow traffic to be allocated to each available interface.

Consequently, the action space size of each head is calculated as the number of permutations of the possible fractions that sum to one. The previous number is obtained using the well-known combinatronics ``stars and bars'' technique\cite{Krickeberg1969}. The size of each head can be calculated as:

\begin{equation}
|A|_{h2} = \frac{(n+1)!}{n!},
\end{equation}

\begin{equation}
  |A|_{h3} = \frac{(n+2)!}{2!n!},  
\end{equation}
where $n=10$. The previous $n$ value is chosen based on the discretization of the maximum flow traffic distribution value using a $0.1$ interval. After the substitution of $n$, we obtain $|A|_{h2} = 11$ and  $|A|_{h3} = 66$.

\subsection{Reward function}
The reward function in the t-$th$ episode is defined as:
\begin{equation}
    R_t = 1 - D_t^{avg},
\end{equation}

where $ D_i^{avg}$ corresponds to the average throughput drop ratio observed by the $m^{th}$ AP. In addition, we scale the reward to $[-1,1]$.
\subsection{Reward function with hindsight}
In \cite{Andrychowicz2017}, the authors present a buffer technique called \textit{Hindsight Experience Replay} (HER) that allows the successful application of model-free RL algorithms in environments characterized by sparse rewards. The idea involves the usage of \textit{goals states} to improve the sample efficiency and convergence. Differently from the previous work, we propose to utilize goals not in the state space but in the reward function itself. Thus, we define the term \textit{goal reward} as the threshold in which given a reward from the environment a penalization is applied to the composite reward function as described as follows:
\begin{equation}
     R_t^h =
    \begin{cases}
     R_t & \text{if }  D_t^{avg} < D_{TOL},\\
      -1 & \textit{otherwise}
    \end{cases}
    \label{intrinsic}
\end{equation}

where $D_{TOL}$ corresponds to a hindsight reward based on baselines results or on expert knowledge. The intuition behind the before mentioned formulation relays on the more number of times the agent obtains a reward worse than the \textit{goal reward} given an action $\bm{a}$, the stronger the indication that the action taken was not positive given an observed state. Eventually, the agent will be able to learn from those undesired actions and start learning from the desired ones.

\subsection{Complexity Analysis }

The complexity of our proposed MH-RSAC scheme corresponds to the complexity of the neural network comprising the SAC base algorithm. Thus, the training complexity is $O( \mathcal{H} \cdot \mathcal{W} \cdot \frac{|\mathcal{D}|}{B} \cdot T)$, where $\mathcal{H}$ represents the learning update periodicity, $\mathcal{W}$ the number of updates per session, $\mathcal{D}$ the buffer size, $B$ the batch size and $O(T)$ denotes the time complexity of a single iteration.

\subsection{Baselines: SLCI and MCAA}
In this work, we compare our proposed scheme with two non-RL approaches described in \cite{Lopez-Raventos2021}.

\textbf{Single Link Less Congested Interface (SLCI):} The traffic is allocated to the less congested interface. 

\textbf{Multi-Link Congestion-aware Load balancing at flow arrivals (MCAA):} The traffic is distributed among the available interfaces given the observed occupancy per interface at the AP.

Note that, in a recent letter \cite{9765523} the authors have presented a dynamic traffic allocation proposal. However, different than the baselines and our proposed approach, such policy could modify the flow distribution over the available interfaces periodically without the constrain of only doing so upon an incoming flow.

\section{Traffic considerations} \label{section5}
As previously mentioned, IEEE 802.11be aims to respond to the increasing demand of high-throughput services. Among them, we encounter Virtual Reality (VR). In \cite{9685808}, the authors present an empirical model based on two campaigns of VR gaming data measurements. The first campaign is run using a VR game (``Beat Saber'') in a local server and the second in a cloud server, respectively. In both cases, the frame size and frame inter-arrival time are measured using Wireshark. Results showed that the frame size behaved as a loglogistic distribution and the frame inter-arrival time as a Burr distribution. In this work, we utilize the cloud server model and derive the Cumulative Distribution Functions (CDF) to be utilized as one of our simulation traffic flows. The derivation of the inverse of the CDF for the frame inter-arrival time and frame size is described below: 
\begin{figure}
\center
  \includegraphics[scale=0.8]{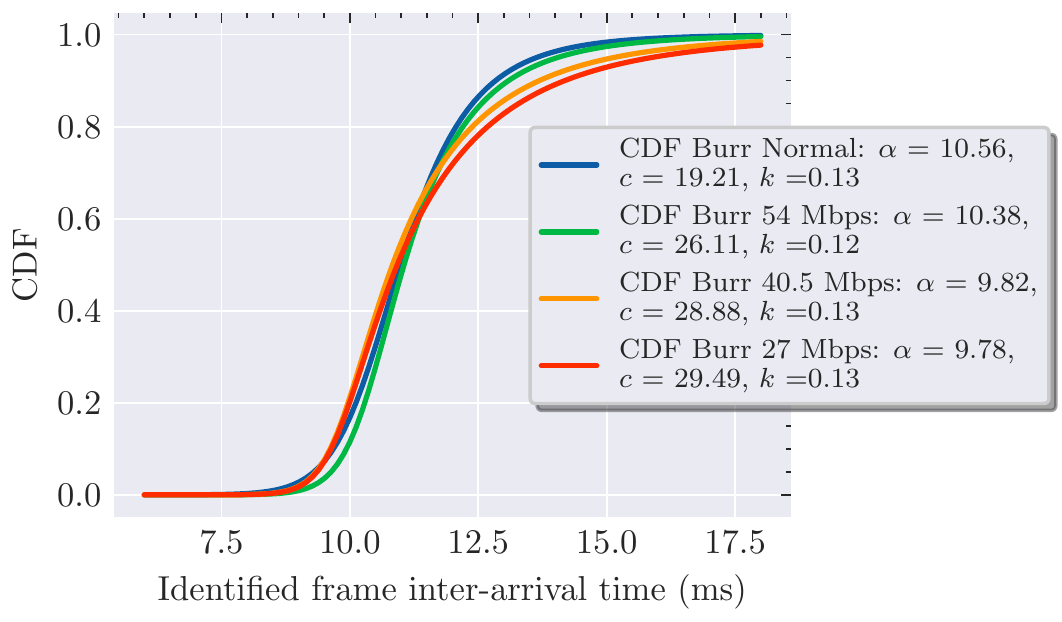}
  \caption{Identified frame inter-arrival time CDF}
 \label{burr_cdf}
\end{figure}

\begin{figure}

  \includegraphics[scale=0.8]{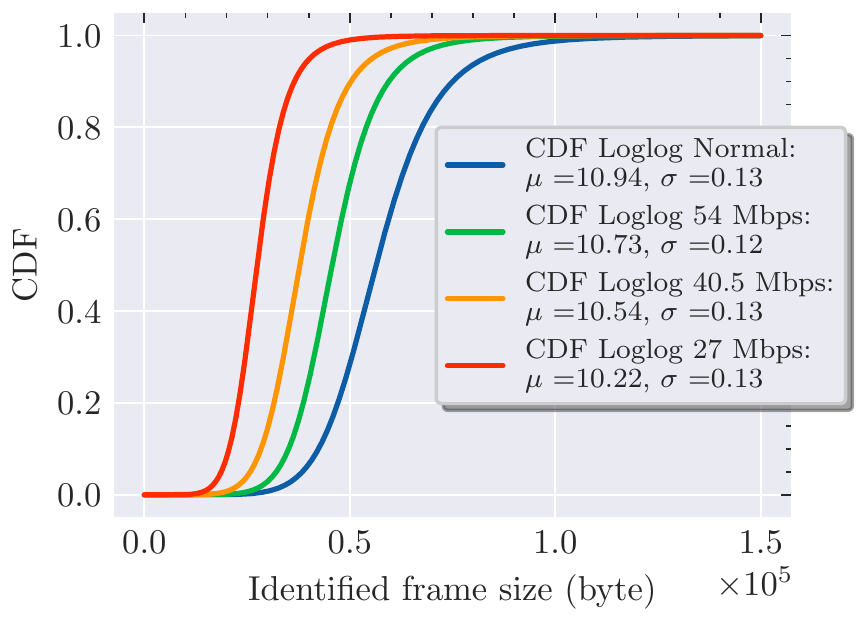}
  \caption{Identified frame size CDF }
 \label{loglog_cdf}
\end{figure}

\begin{equation}
   F_{fs}(x) = \int_{0}^{\infty} \frac{e^{\frac{\ln(x)-\mu}{\sigma}}}{\sigma x (1 + e^{\frac{\ln{x} - \mu}{\sigma}})^2} \,dx  = - \frac{1}{e^{\frac{\ln(x)-\mu}{\sigma}} + 1} + C,
\end{equation}

\begin{figure*}
\center
  \includegraphics[scale=0.7]{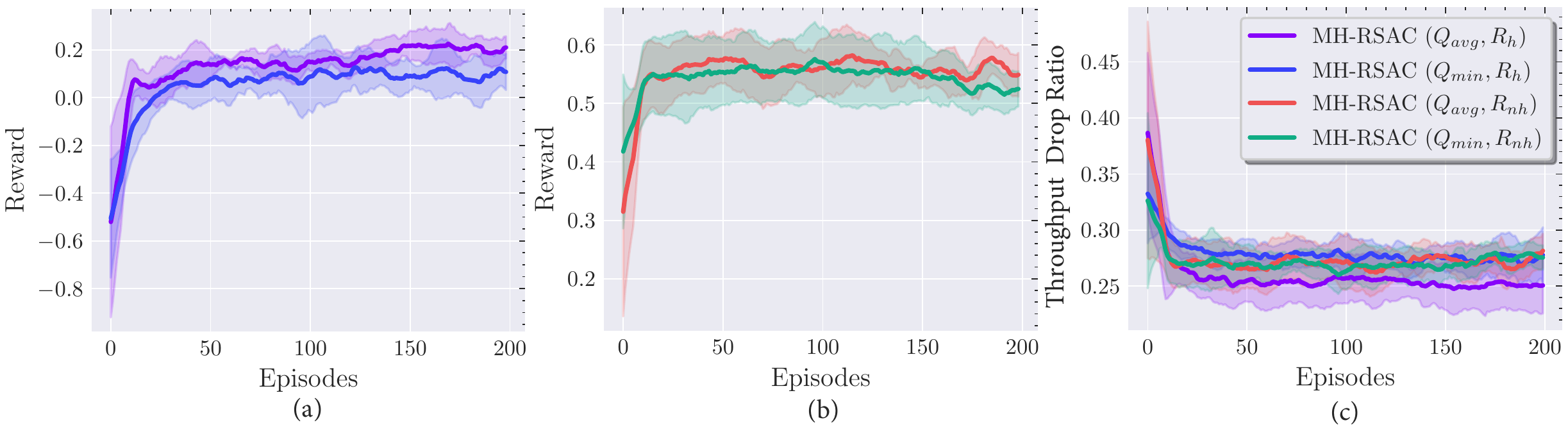}
  \caption{Reward and TDR convergence for the ``\textbf{avg}'', ``\textbf{min}'' operator of the MH-RSAC's variants in the \textbf{U2} scenario: \textbf{(a)} Reward with hindsight $(R_h)$, \textbf{(b)} Reward without hindsight  $(R_{nh})$ and \textbf{(c)} TDR convergence.}
 \label{drop_convergence}
\end{figure*} 

where $F_{fs}$ corresponds to the frame size CDF, $\mu$ and $\sigma$ are the statistical model parameters of the loglogistic distribution and $C$ a constant related to the mathematical integration. Finally, the inverse of the CDF in bytes is: 
\begin{equation}
 F_{fs}^{-1}(y) = e^{\mu}(-\frac{y}{y-1})^\sigma [bytes],
\end{equation}
The frame inter-arrival time can be described as follows:
\begin{equation}
F_{ft}(x) = \int_{0}^{\infty} \frac{\frac{kc}{a}(\frac{x}{a})^{c-1}}{\left(1+ (\frac{x}{a})^c \right)^{k+1}} \,dx  = - \frac{1}{\left((\frac{x}{a})^c + 1\right)^k} + C,
\end{equation}
where $F_{ft}$ corresponds to the frame inter-arrival time CDF, $k$, $a$ and $c$ are the statistical model parameters of the Burr distribution. The inverse of the CDF in ms is:
\begin{equation}
 F_{ft}^{-1}(y) =  a\sqrt[\leftroot{1} \uproot{5} c]{\sqrt[\leftroot{1} \uproot{5} k]{\frac{1}{1-y}} - 1} [ms],
\end{equation}
Finally the VR flow size in Mbps can be calculated as: 
\begin{equation}
 f_{VR}(y) =  \frac{F_{fs}^{-1}(y)}{F_{ft}^{-1}(y)} [Mbps], 
\end{equation}
where $y$ is randomly sampled as $y\sim{\mathcal{U}(0,1]}$. In Fig. \ref{burr_cdf} and  \ref{loglog_cdf}, we present the VR CDFs for the frame inter-arrival and frame size considering AP bandwidth throttling of $54$ Mbps, $40.5$ Mbps and $27$ Mbps and no AP throttling (Normal), respectively. Note that only downlink traffic is considered and all traffic flows are modeled as a single Constant Bit Ratio (CBR) flow. For this work purpose the VR traffic generated corresponds to the non-throttling case (Normal). 
Besides the VR traffic, we assume the coexistence of two more traffic flows. One depicting an HD/4K flow with a size of $f_{4K} \sim {\mathcal{U}(7,25)}$ Mbps and a web browsing traffic with a size of $f_{WB} \sim {\mathcal{U}(1,3)}$ Mbps. At the beginning of each simulation, the data flow types are distributed among the stations attached to their AP as follows: \{WB: $0.8$, V4K: $0.1$, VR: $0.1$\}.

\section{Performance evaluation}\label{secton6}

Simulations are performed using the flow-level simulator Neko 802.11be\cite{Lopez-Raventos2021} and Pytorch-based RL agents. The communication between the simulator and the RL agents is done using the ZMQ broker library.


\subsection{Simulation Settings}
Simulation settings and RL parameters utilized in this work are depicted in Table \ref{net_settings} and \ref{learning_settings}, respectively. Furthermore, we consider two load distribution scenarios: \textbf{U1} with number of users attached to its corresponding AP, $N_A \sim \mathcal{U}(15,20)$ and \textbf{U2} with $N_A \sim \mathcal{U}(20,25)$. In the next subsection we will discuss the performance results of the proposed scheme.

\begin{table}
\caption{Network Settings}
\begin{center}
\resizebox{\columnwidth}{!}{%

\begin{tabular}{c c} 
\hline
\textbf{Parameter}&\textbf{Value} \\
\hline
IEEE protocol & { 802.11be } \\
Channel Bandwidth & { $20$ MHz/$40$ MHz/$80$ MHz/$160$ MHz } \\
Carrier Frequency ($f_c$) & { $2.437$ GHz/$5.230$ GHz/$6.295$ GHz } \\
Max Modulation/Max Modulation Coding Scheme & { 4096 QAM/MCS 13}\\
Number of APs & { 5 } \\
Number of Stations per AP & {  \textbf{U1} : $N_A \sim \mathcal{U}(15,20)$,  \textbf{U2} : $N_A \sim \mathcal{U}(20,25)$}\\
Max Spatial Streams & { 16 } \\ 
Propagation Loss Model & { $P_{l}(d) = 40.05 + 20\text{log}(f_{c}/2.4) + 20\text{log}(\text{min}(d,10)) +$} \\ {} & {$(d>10) * 35\text{log}(d/10) + 7W$} \\
Data Flow Types & { Web browsing (WB), Video 4K (V4K), Virtual Reality (VR) }
\\
Data Flow User Distribution & { WB: $0.8$, V4K: $0.1$, VR: $0.1$ }
\\
AP/STA Noise Figure & {$7$ dB}\\
AP/STA Transmission Power & {$20$/$15$ dBm}\\

CCA threshold & {-82 dBm}\\
Packet Error Rate & {$10$\%}\\

\hline
\end{tabular}
}
\label{net_settings}

\end{center}
\end{table}

\begin{figure*}
\centering

  \includegraphics[scale=0.75]{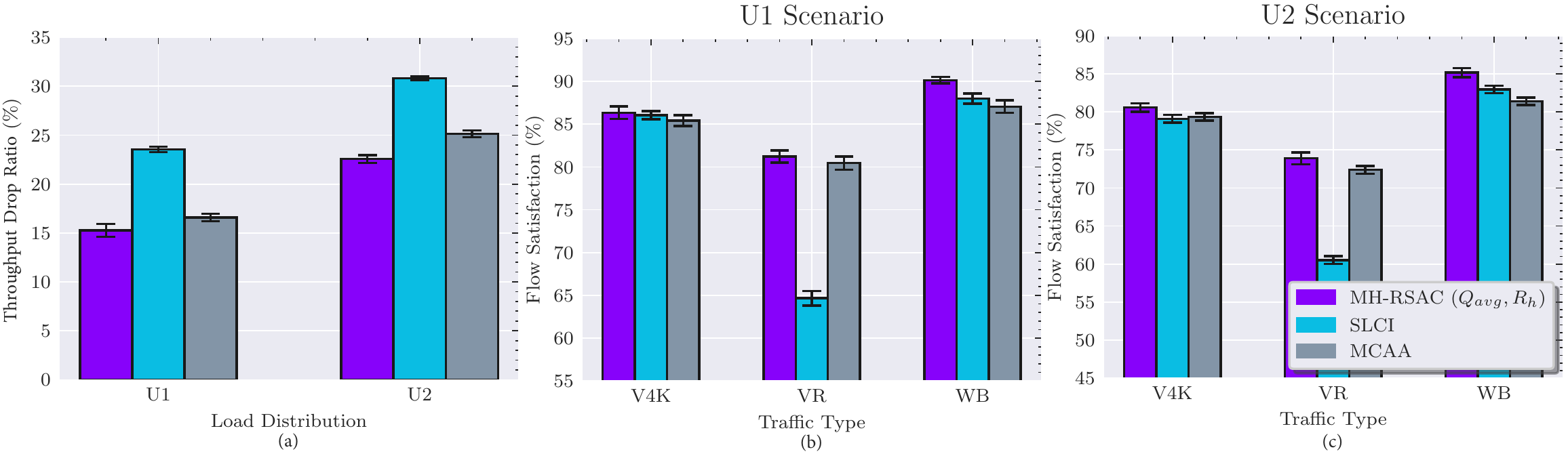}
  \caption{\textbf{(a)} Throughput Drop Ratio (TDR) per load distribution \textbf{U1} and \textbf{U2} and Flow Satisfaction (FS) per traffic for Video HD/4K (V4K), Virtual Reality (VR) and Web Browsing (WB) per load distribution \textbf{(b)} \textbf{U1}  and \textbf{(c)} \textbf{U2}   }
 \label{flow_sat_drop}
\end{figure*}

\subsection{Simulation Results}
We present the performance results of our proposed scheme in terms of convergence, FS per interface and TDR.  Figure \ref{drop_convergence} shows the convergence behavior of the \textbf{U2} scenario for the SAC variants proposed in this work. The figures that depict the scenario \textbf{U1} are not shown in this work, as they show similar information. Figure \ref{drop_convergence} (a) shows how the ``\textbf{avg}'' operator outperforms the ``\textbf{min}'' operator when the reward with hindsight is considered. Conversely, the previous behavior does not repeat when the rewards without hindsight are utilized as shown in Fig. \ref{drop_convergence} (b). Finally, we show the convergence in terms of TDR for the four variants in Fig. \ref{drop_convergence} (c) where the best convergence is achieved when the ``\textbf{avg}''operator with rewards with hindsight are used. The previous results confirm that the \textit{goal rewards} strategy proposed in this work affects positively to the learning process in non-Markovian scenarios.

\begin{table}
\caption{Reinforcement Learning Settings }
\tiny
\begin{center}

\begin{tabular}{c c} 
\hline
\textbf{Parameter}&\textbf{Value} \\
\hline

Actor learning rate & { $1\mathrm{e}{-4}$ } \\
Actor final layer activation function & {Softmax}\\
Critic learning rate & { $1\mathrm{e}{-3}$ } \\
Critic final layer activation function & {None}\\
Linear hidden layers & { $[64, 64]$ } \\
Window ($W$) & 10 \\ 
Tau ($\tau$) & { $5\mathrm{e}{-3}$ } \\
Discount rate & {$0.99$} \\
Gradient clipping norm & {$1$ }\\
Weight initializer & { Xavier } \\ 
Batch size & { $512$ }\\
Learning updates periodicity ($\mathcal{H}$) & { $50$ steps }\\
Learning updates per learning session  ($\mathcal{W}$) & { $10$ }\\
Automatic entropy tuning & {Enabled}\\

\hline
\end{tabular}

\label{learning_settings}

\end{center}
\end{table}
Furthermore, in Fig. \ref{flow_sat_drop} we present the results in terms of TDR and FS per traffic type for \textbf{U1} and \textbf{U2} scenarios, respectively. In Fig. \ref{flow_sat_drop} (a), we show the TDR performance of the MH-RSAC scheme with ``\textbf{avg}'' operator and rewards with hindsight and the two baselines: SLCI and MCAA. Consequently, we observe that the MH-RSAC scheme offers a gain of $34.2\%$ and $35.2\%$ when compared with the SLCI algorithm and $2.5\%$ and $6\%$ when compared with the MCAA algorithm for the \textbf{U1} and \textbf{U2} scenarios, respectively. Additionally, we present a comparison in terms of FS per traffic type in Fig. \ref{flow_sat_drop}(b) and (c). For instance, in Fig. \ref{flow_sat_drop} (b) we can see that the MH-RSAC scheme's gains for the V4K traffic around $1\%$ but when it comes to the VR traffic which is a more throughput hungry service SLCI is not capable to perform positively with a reduction of $25.6\%$ in terms of FS. Moreover, for the dynamic WB traffic the MH-RSAC scheme shows an improvement of $2.5\%$ and $4\%$ over the SLCI and MCAA algorithms. In a similar fashion in Fig. \ref{flow_sat_drop} (c),  we obtain a gain of $21\%$ and $3\%$ for the VR traffic and $3.2\%$ and $6\%$ for the WB traffic of the MH-RSAC algorithm over the two baselines SLCI and MCAA, respectively. To sum, we can see that MH-RSAC is capable of distributing high throughout traffic over multiple links and also respond efficiently to more dynamic traffic types like Web Browsing.

\section{Conclusions } \label{Section6}

In this paper, we presented a Soft-Actor Critic (SAC) Reinforcement Learning (RL) agent to improve performance metrics such as Throughput Drop Ratio (TDR) and Flow Satisfaction (FS) in IEEE 802.11be MLO capable networks. More specifically, we proposed an agent named Multi-Headed Recurrent Soft-Actor Critic (MH-RSAC) that is capable to deal with different types of Multi-Link Devices (MLD)s. We included the modifications required in its design such as the ``\textbf{avg}'' operator to reduce underestimation in the SAC algorithm. In addition, we acknowledged the non-Markovian nature of the scenario under study and included two main techniques that involve the usage of Long Short-Term Memory (LSTM) neural networks and rewards with hindsight. Moreover, we presented the derivation of a Virtual Reality (VR) traffic to be further utilized in any wireless simulator. We compared the proposed RL variants in terms of convergence and observed the best performance for the case of the MH-RSAC with  ``\textbf{avg}'' operator and utilization of rewards with hindsight (MH-RSAC$(Q_{avg}, R_h)$). Furthermore, we presented the simulation results of our proposed scheme and two previously defined baselines named \textit{Single Link Less Congested Interface} (SLCI) and \textit{Multi Link Congestion-aware Load balancing at flow arrivals} in terms of TDR and FS per traffic type. Results showed that MH-RSAC$(Q_{avg}, R_h)$ outperforms in terms of TDR the SLCI baseline with an average gain of $34.2\%$ and $35.2\%$ and the MCAA baseline with $2.5\%$ and $6\%$ in the \textbf{U1} and \textbf{U2} proposed scenarios, respectively. Finally, we observed that our scheme is able to respond more efficiently to high throughput and dynamic traffic such as VR and Web Browsing (WB) when compared with the baselines. Results showed an improvement of the MH-RSAC scheme in terms of FS of up to $25.6\%$ and $6\%$ for the SCLI and MCAA, respectively. As future work we intend to study with more details how to tackle more effectively non-Markovian scenarios in wireless networks and more specifically in IEEE 802.11be MLO capable networks.

\section{Acknowledgment }\label{Section7}
This research is supported by Mitacs Accelerate Program and NetExperience Inc.


\bibliography{biblio.bib}{}
\bibliographystyle{IEEEtran}

\end{document}